\documentclass[conference]{IEEEtran}
\IEEEoverridecommandlockouts

\usepackage{setspace}
\usepackage{amssymb}
\usepackage{paralist}
\usepackage{amsmath}
\usepackage{amsthm}
\usepackage{graphicx}
\usepackage{epstopdf}
\usepackage{subfig}
\usepackage{color}
\usepackage{extarrows}
\usepackage{multicol}
\usepackage{caption}
\usepackage{cite}
\usepackage{url}
\usepackage{bm}
\usepackage{footnote}
\usepackage{multirow}
\usepackage{array}
\usepackage{diagbox}
\usepackage[lined,ruled,boxed,linesnumbered]{algorithm2e}
\usepackage[table,xcdraw]{xcolor}

\allowdisplaybreaks

\begin{document}
\setlength{\parskip}{-0.5pt}
\title{An SDN-Based Transmission Protocol with In-Path Packet Caching and Retransmission
}	
	\author{
		\IEEEauthorblockN{Jiayin Chen\IEEEauthorrefmark{1}, Si Yan\IEEEauthorrefmark{1}, Qiang Ye\IEEEauthorrefmark{1},Wei Quan\IEEEauthorrefmark{2}, Phu Thinh Do\IEEEauthorrefmark{1}, Weihua Zhuang\IEEEauthorrefmark{1}, Xuemin (Sherman) Shen\IEEEauthorrefmark{1} \\ Xu Li\IEEEauthorrefmark{3}, Jaya Rao\IEEEauthorrefmark{3}}
		
		\IEEEauthorblockA{\IEEEauthorrefmark{1}Department of Electrical and Computer Engineering, University of Waterloo, Waterloo, ON, N2L 3G1,  Canada\\\{j648chen,s52yan,q6ye,pt3do,wzhuang,sshen\}@uwaterloo.ca}
		\IEEEauthorblockA{ \IEEEauthorrefmark{2}School of Electronic and Information Engineering, Beijing Jiaotong University, Beijing, 100044, P.R. China\\dr.wei.quan@ieee.org}
		\IEEEauthorblockA{ \IEEEauthorrefmark{3}Huawei Technologies Canada Inc., Ottawa, ON, K2K 3J1, Canada\\\{xu.lica,jaya.rao\}@huawei.com}	
	}
\maketitle

\begin{abstract}
In this paper, a comprehensive software-defined networking (SDN) based transmission protocol (SDTP) is presented for fifth generation (5G) communication networks, where an SDN controller gathers network state information from the physical network to improve data transmission efficiency between end hosts, with in-path packet retransmission. In the SDTP, we first develop a new two-way handshake mechanism for connection establishment between a pair of end host. With the aid of SDN control module, signaling exchanges for establishing E2E connections are migrated to the control plane to improve resource utilization in the data plane. A new SDTP packet header format is designed to support efficient data transmission with in-path packet caching and packet retransmission. Based on the new data packet format, a novel in-path receiver-based packet loss detection and caching-based packet retransmission scheme is proposed to achieve in-path fast recovery of lost packets.
Extensive simulation results are presented to validate the effectiveness of the proposed protocol in terms of low connection establishment delay and low end-to-end packet transmission delay.
\end{abstract}


\section{Introduction}
\label{sec:Introduction}
With fierce advancement of networking technologies, the fifth generation (5G) communication networks are foreseen to accommodate diversified bandwidth-hungry applications with stringent quality-of-service (QoS) requirements. Because of current distributed and ossified core network architecture with limited computing and transmission resources on network servers/switches and links, differentiated end-to-end (E2E) QoS guarantee for various services is always difficult to achieve \cite{parvez2018survey}. It is imperative to develop an efficient transport-layer protocol to reduce network congestion and enhance E2E QoS satisfaction. Transmission control protocol (TCP) \cite{postel2003rfc} is a typical transport-layer protocol widely used in the Internet to achieve reliable end-to-end packet transmissions. In TCP, through a three-way handshake procedure during the connection establishment phase, a pair of end hosts establish a two-way communication connection for data transmission. The sending host uses retransmission timeout (RTO) and fast retransmission with congestion window adjustment for lost packet recovery and congestion control. Different TCP variants \cite{chaudhary2017review} are developed to improve the E2E performance under different network scenarios (e.g., a network with high bandwidth delay product).

To support a high data traffic volume and different levels of E2E delay requirements from diversified services, more and more network elements (i.e., servers and switches) are placed into the network. Due to highly dynamic traffic load, resources at some network locations are underutilized, whereas other network elements may experience high traffic congestion and loss. There exist studies on how to improve the loss recovery performance of transport-layer protocols by reducing the probability of false fast retransmissions and improving the accuracy of RTO estimation \cite{globisch2014retransmission}. However, current transport-layer protocols, under existing network architecture, e.g., TCP and user datagram potocol (UDP) \cite{tinta2009characterizing}, only achieve best-effort E2E performance, due to slow reaction to packet loss. Software-defined networking (SDN) \cite{nunes2014survey, gu2017real, han2017sd} emerges as a promising network architecture to achieve more fine-grained in-network control. SDN decouples control functions from network servers/switches and migrates them as an integrated and centralized control module, which simplifies packet forwarding functions and makes loss recovery decisions on individual servers/switches.

Current transport-layer protocols rely on end hosts to detect packet loss and perform packet retransmission. 
With SDN, it is essential to investigate how transport-layer protocols can be enhanced to perform prompt packet loss (or congestion) detection and recovery. Existing studies exploit the SDN control module to gather in-network statistics (e.g., buffer occupancy) from OpenFlow switches for early packet loss (or congestion) detection and fast response \cite{hafeez2018detection}.
In \cite{wang2017sdudp}, an SDN-based UDP framework is proposed, where UDP is employed for transmission between a pair of edge switches to reduce communication overhead, and a retransmission engine is activated on each network switch to perform in-path packet loss detection and packet retransmissions. Packet loss is detected simply based on out-of-order packets observed at an in-path switch, without a retransmission policy to distinguish between packet loss and packet reordering. Moreover, in most of existing works\cite{wang2018r,chen2011reliable}, it is assumed that every switch is enabled caching and retransmission function, which inevitably incurs substantial resource consumption for caching at each switch and signalling overhead between switches for retransmission request and caching release. In this paper, we develop a comprehensive SDN-based transmission protocol (SDTP) with the consideration of in-path packet caching and retransmission. The protocol framework consists of following three parts, of which the main contributions are summarized:
\begin{enumerate}[(1)]
\item \emph{Connection establishment} -- We propose an SDN-based two-way connection establishment procedure for each pair of end hosts. With SDN controller, signaling exchanges for establishing each E2E connection are migrated to the control plane to save resources for data transmissions. Compared with the TCP three-way handshake, the proposed procedure reduces the amount of signaling exchanges overhead for establishing bi-directional connections;

\item \emph{Data transmission} -- We design a new header format to support efficient data transmission, in-path caching and retransmission. On one hand, unnecessary fields in the TCP/IP header are removed because the controller can offload partial functions from switches in data plane. On the other hand, we introduce new fields to support advanced protocol function elements, such as packet caching and receiver-triggered packet retransmission, which is different from the conventional TCP;

\item \emph{Caching-based packet retransmission} -- A novel in-path receiver-based packet loss detection and caching-based packet retransmission scheme is proposed to achieve in-path congestion detection and fast recovery of lost packets with low signaling overhead, which further reduces E2E packet transmission delay especially when congestion happens. The receiver-triggered packet retransmission is promising in terms of short response delay in packet loss detection and in reducing state management complexity at sender sides \cite{oh2012new}. 

\end{enumerate}

\section{System Model}
\label{sec:System_Model}
The system model under consideration includes an embedded virtual network topology, a description of caching and retransmission function elements and their placement policies, and a traffic model for each end host.
\vspace{-0.1cm}
\subsection{Embedded Virtual Network Topology}
\label{sec:SM_NM_NT}
\begin{figure}[!htb]
	\vspace{-0.2cm}  
	\setlength{\belowcaptionskip}{-0.4cm}   
	\centering
	{\includegraphics[width=0.85\linewidth]{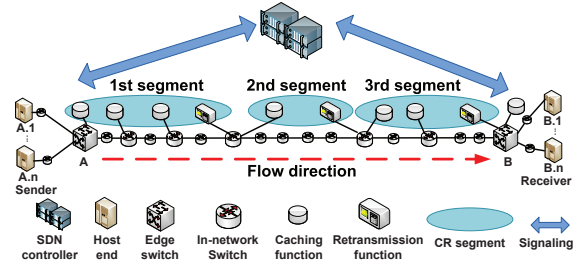}}
	\caption{An embedded network topology for different E2E connections.}\label{fig:network topology}
\end{figure}

Given a service description and its QoS requirement, our SDT algorithm \cite{Omar2018vnf} determines an embedded topology for each virtual network function (VNF) chain on the physical substrate network. Based on user requests, every source-destination (S-D) host pair has an E2E data connection for reliable transmissions, in which the routing path between two edge switches for each connection is established through the SDN controller caching flow entries (containing forwarding rules) into each network switch according to the SDT output.  After one E2E connection is established, data packets are sent from the source host (or the sending host), and are then forwarded via a pair of edge switches and other network switches and transmission links to reach the destination host (or the receiving host). For example, in Fig.~\ref{fig:network topology}, there are in total $n$ hosts connected to edge switch $A$, denoted by $\{A.1, A.2, ..., A.n\}$, which are supposed to establish connections with the set of hosts $\{B.1, B.2, ..., B.n\}$ under $B$, respectively. It is assumed that all end hosts under one edge switch belong to one service type and the traffic flow\setlength{\footnotesep}{0.2cm}\footnote{A traffic (service) flow refers to an aggregation of packets of same service type passing through a pair of edge switches in the core network.} aggregated at the edge switch are forwarded via a single-path routing in the core network to the other edge switch. 
\vspace{-0.2cm}
\vspace{-0.1cm}
\subsection{Caching and Retransmission Function Elements}
\label{sec:SM_NM_FUN}

To facilitate packet loss detection and packet retransmission, we activate caching and retransmission function elements at in-path switches.

\emph{Caching Function Element} -- A caching node is defined as a network switch with the caching function activated. Each caching node is equipped with pairs of data transmission and caching buffers. Each data buffer is used to queue the data packets for packet forwarding. After a packet is processed by the node, it is cached (i.e. copied) in its caching buffer. 
Each cached packet is used for in-network early retransmission if the packet is lost on a network path, instead of being retransmitted from the sender side.
 Each connection has a unique pair of data transmission and caching buffers. Note that data buffer overflow leads to data packet loss\setlength{\footnotesep}{0.2cm}\footnote{Note that packet loss refers to data packet loss in this paper.}, while caching buffer overflow causes cached packet loss.
Thus, buffer space release is necessary to avoid caching buffer overflow, which is realized by periodically transmitting a caching notification (CN) packet upstream from one caching node to its previous one for releasing cached packets.

\emph{Retransmission Function Element} -- A retransmission node refers to an in-path switch with the functionalities of in-path packet loss detection and retransmission triggering. After packet loss is detected at a retransmission node, a retransmission request is sent from the retransmission node to each preceding upstream caching node consecutively until the requested data packets are located. We define one \emph{caching-retransmission (CR) segment} as a network segment including one retransmission node, network switches/transmission links, and all the caching nodes between this retransmission node and its nearest upstream retransmission node. 
\vspace{-0.1cm}
\subsection{Caching and Retransmission Placement}
\label{sec:SM_NM_FUNPL}
For each E2E host pair, a basic policy is that the sending edge switch is equipped with the caching function element, while the receiving edge switch enables both caching and retransmission function elements to minimize the maximum packet retransmission delay. Packet retransmission delay is the time duration from the instant that packet loss is detected and a retransmission request is sent to the instant that a retransmitted packet is received by the retransmission node. In addition, to ensure packet caching between consecutive CR segments, the retransmission node of one CR segment is one of the caching nodes of its subsequent CR segment. An example of caching and retransmission functionality placement for one pair of edge switches is shown in Fig.~\ref{fig:network topology}.
The policy of caching and retransmission function placement is based on packet loss probability of each link. Caching functions are activated at the switch nodes before links with high packet loss probabilities, and then some of these caching nodes are selected for retransmission function activation. To balance resource utilization of different CR segments, we choose retransmission nodes so that the packet loss probabilities over each segment are approximately equalized, named the \emph{equalized loss probability (EP) policy}. This policy indicates the chances of packet loss over each segment are similar, and thus the numbers of retransmission requests sent over each segment are balanced.


\section{Software-Defined Transmission Protocol}
\label{sec:SDP_operation}
\subsection{Connection Establishment}
\label{sec:SDP_CE}
\begin{figure}[!htbp]
	\centering
	\setlength{\belowcaptionskip}{-0.4cm}   
	{\includegraphics[scale=0.26]{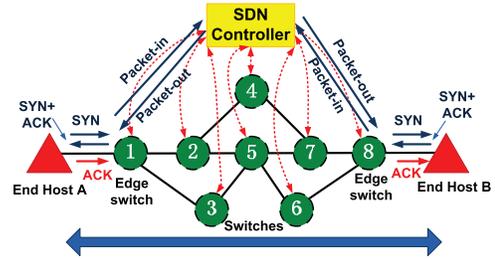}}
	\caption{The SDTP signaling during the connection establishment phase.}\label{figs:connection_establishment_signals}
\end{figure}

An SDN controller has a global view over the network states and can check edge-to-edge path availability, and it also has the capability to set up a routing path, including activating caching and retransmission functionalities during the connection establishment phase to satisfy E2E service requirement. With the SDN control module, we can adopt a two-way handshake, instead of the TCP three-way handshake, for connection establishment in SDTP to reduce the signaling overhead, shown in Fig.~\ref{figs:connection_establishment_signals}: (1) Sending edge switch 1 receives a synchronous (SYN) packet sent by end host A; (2) Edge switch 1 encapsulates the SYN packet into an OpenFlow packet (i.e., Packet-in) and sends it to the SDN controller. This process can be implemented by configuring a new flow entry specific for SYN packets in the edge switch. This flow entry points to an encapsulation action; (3) The controller receives and parses the OpenFlow packet, and then checks the path reachability according to a table of local link status. If there is a reachable path, the controller continues to forward this packet to receiving edge switch 8 through another OpenFlow packet (Packet-out), and then the receiving edge switch abstracts the pure SYN packet and sends it to end host B. If there is no available path, the controller deploys a new path. The path calculation is based on the SDT algorithm \cite{Omar2018vnf}; (4) A similar process is conducted for sending back the SYN-Acknowledgement (ACK) packet, where the SDN controller checks the reachability for establishing a reverse connection; (5) To be compatible with conventional TCP employed by end hosts, we make some patch processing. That is to notify end host B, by an additional ACK packet, that the connection establishment is accomplished. Note that the ACK packet does not come from end host A, but is triggered by the receiving edge switch. At the same time, the sending edge switch requires to drop the ACK packet received from end host A directly.
\vspace{-0.3cm}

\subsection{Data Transmission}
\label{sec:SDP_DT}
\begin{figure}[!htbp]
	\centering
	\setlength{\belowcaptionskip}{-0.3cm}   
	{\includegraphics[width=0.46\textwidth]{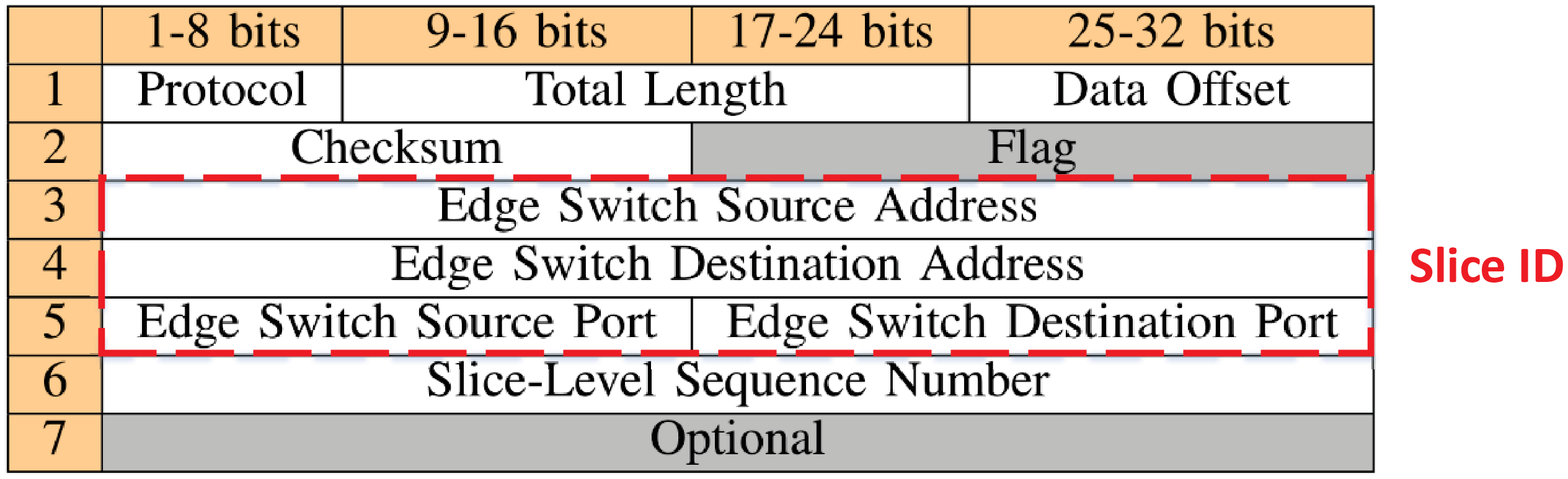}}
	\caption{The SDTP packet header formats.}\label{figs:SDATP_M2M_packet_header_formats}
\end{figure}

For compatibility, the packet transmission between an end host and an edge switch is based on conventional TCP, whereas the transmission between edge switches follow our proposed SDTP protocol. Therefore, the sending and receiving edge switches should execute header conversion and reversion for E2E communication.
In SDTP, a new packet header format is developed to support efficient and reliable data transmission as shown in Fig. \ref{figs:SDATP_M2M_packet_header_formats}, which includes $24$-byte required fields and $20$-byte optional field. On one hand, some fields in the TCP/IP header are removed since the controller can offload partial functions from switches in the data plane. For example, the Acknowledgment field in the conventional TCP header is removed from the SDTP data packet because SDTP adopts the receiver-triggered packet loss detection. In this way, the SDTP data transmission achieves higher throughput according to the simplified packet header. On the other hand, two new fields, Flag and Optional fields, are added into the SDTP header for new functionalities different from the conventional TCP, such as in-path caching and caching-based retransmission.
\vspace{-0.2cm}

The header supports OpenFlow-switches, in which some required fields (i.e., slice ID) are extracted to match the flow table entries for packet forwarding along the embedded SDT routing path. Different types of SDTP packets  are identified by the field of Flag, including retransmission request (RR) packet, retransmission data (RD) packet, retransmission information (RI) packet, and caching notification (CN) packet. The optional field contains different information for different types.
\vspace{-0.1cm}
\subsection{Caching-Based Packet Retransmission}
\label{sec:SDP_RP}
\subsubsection{Terminology for Packet Loss Detection}
\label{sec:SDP_RP_tab}
After the connection establishment for a pair of E2E hosts, a \emph{content window list} and an \emph{expected packet list} are maintained at each retransmission node for packet loss recovery. We elaborate how a content window list and an expected packet list are built up, and define some important variables and parameters independently used by each connection to detect packet loss.

\emph{Content Window List} -- A content window list is established to record the packets already received by a retransmission node. A number of packets received in sequence can be described by one window, where the \emph{left edge} of the window indicates the first sequence number of the received packets and the \emph{right edge} of the window is the next expected sequence number of the last received packet. When packets are not received in sequence, several content windows are generated, and each entry in the list represents a single window. The content window list is updated when a new packet is received at the retransmission node.

\emph{Expected Packet List} -- An expected packet list is used to record the information of packets that are expected by a retransmission node. After packet loss is detected and a retransmission is triggered, the node refers to the expected packet list to sort out the lost packets to be retransmitted. The expected packet list has the following information fields:

\begin{enumerate}[(1)]
	\item \emph{Num} -- When packets are lost discontinuously, different portions of expected packets are inserted into the expected list as different entries. The Num field records the sequence of the entries;
	
	\item \emph{StartSeq} and \emph{EndSeq} -- Specifying an interval with a start sequence number and an end sequence number;
	
	\item \emph{StartNum} -- A pointer indicating the number of packet offset from the packet with the sequence number equaling StartSeq, with initial value 0;
	
	\item \emph{InterCnt} -- A packet interarrival counter indicating the number of packets received after the last sequentially received packet, with initial value 0;
	
	\item \emph{CntThres} -- A threshold for InterCnt to detect packet loss, with initial value 1;
	
	\item \emph{WaitLen} -- Measuring the difference between CntThres and InterCnt, with initial value 1;
	
	\item \emph{RTCnt} -- Counting how many retransmission requests of lost packets are sent, with initial value 0;
	
	\item \emph{RTType} -- Defining how packet loss is detected;
	
	\item \emph{RTTimer} -- The time elapses from the instant that a retransmission request is sent to the instant that the retransmitted packet is received.
\end{enumerate}

StartSeq and EndSeq are established based on the right edge of corresponding content window and left edge of its subsequent content window at a retransmission node. If both StartSeq and EndSeq are specified, the packets with the sequence numbers falling between StartSeq and EndSeq are the expected packets; If EndSeq is not determined (set as infinity), we use StartSeq and StartNum to locate a specific expected packet.
The time duration between consecutive packet reception at a retransmission node is interarrival time, denoted as \emph{InterTime} and reset to $0$ whenever a new packet is received. If the recorded InterTime is larger than a threshold, packet loss due to link congestion is detected.
\subsubsection{Thresholds for Packet Loss Detection}
\label{sec:SDP_RP_PL}
We differentiate original packet loss and retransmission packet loss. The original packet loss is triggered by InterTime exceeding a threshold or the number of received disordered packets exceeding a threshold; For retransmission packet loss detection, retransmission RTT is measured for timeout detection.

\emph{Interarrival Timeout} -- We define the threshold for packet interarrival timeout as \emph{expected interarrival time}, which indicates that packet loss is detected when InterTime is greater than the expected interarrival time. The expected interarrival time, denoted by $\Delta T_E$, can be obtained by linear prediction based on sampled interarrival time \cite{begen2004timely},
\begin{equation}
\setlength{\abovedisplayskip}{3pt} 
\setlength{\belowdisplayskip}{3pt}
\Delta T_E = \max\{ \Delta T, a \cdot \Delta T_S + b \cdot \Delta T \}
\label{equ:T_E}
\end{equation}
where $\Delta T_E$ is an estimated expected interarrival time, $\Delta T$ is the sending interval, $\Delta T_S$ is one sample of an interarrival time, $a = 0.875$ and $b = 0.375$.
If interarrival time equals $n \cdot \Delta T_E$ ($n$ is a positive integer), retransmission is triggered by the interarrival timeout. Then, the packet in the expected packet list with smallest WaitLen that has not been retransmitted is selected for retransmission.

\emph{Interarrival Counter Threshold} -- For a single-path routing scenario shown in Fig.~\ref{fig:network topology}, out-of-order packet reception indicates packet loss, and a retransmission node triggers packet retransmission depending on the level of packet disorder. Since the RD packet also leads to disordered packet reception at the following segments, the following retransmission nodes should estimate an updated packet disorder length (CntThres) to avoid spurious packet loss detection. To balance the tradeoff between accuracy and complexity, we determine two-level packet interarrival counter threshold: One is packet-level CntThres which is differentiated for each packet, and the other is segment-level CntThres that is differentiated for each segment.

\begin{figure}[!htbp]
	\centering
	{\includegraphics[scale=0.45]{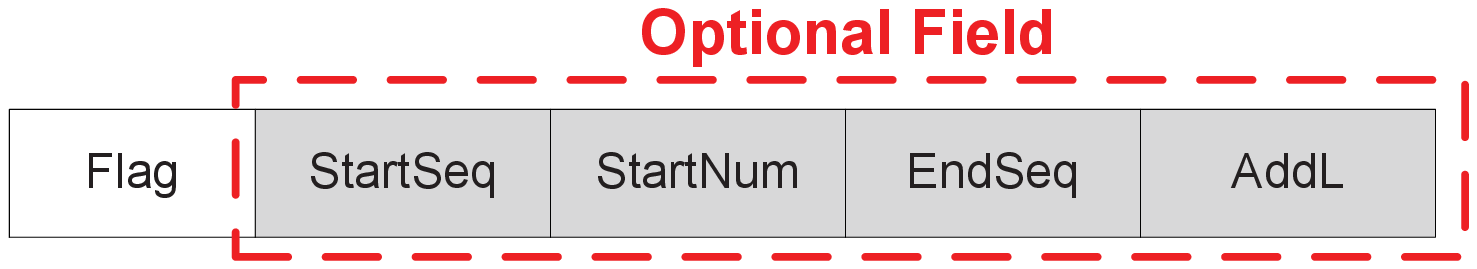}}
	\caption{Main fields in an RI packet.}\label{fig:RI_packet}
\end{figure}

To compute packet-level CntThres, packet retransmission information needs to be shared. If the retransmission node in the $k\,$th CR segment sends an RR packet, the node also sends an RI packet to its following retransmission nodes. The main fields in an RI packet is shown in Fig.~\ref{fig:RI_packet}, in which AddL (i.e., additional packet disorder length) is computed at the $k\,$th retransmission node to estimate how many packets are transmitted ahead of an RD packet. After the following retransmission nodes receive an RI packet, they update CntThres by adding AddL for the retransmitted packet from the $k\,$th CR segment based on received information in the RI packet. 
\vspace{-0.5cm}
\vspace{-0.3cm}
%
If the RR packet is sent at the $k\,$th retransmission node, the additional packet disorder length (AddL), $L_k$, is calculated as
\begin{equation}
\setlength{\abovedisplayskip}{3pt} 
\setlength{\belowdisplayskip}{3pt}
L_k = {R_k}/{I_k} +{a_k}, \, k \in \mathbf{Z^{+}}
\label{equ:Add_Dis_D}
\end{equation}
where $R_k$ is the retransmission RTT and $I_k$ is expected interarrival time at the $k\,$th retransmission node; $a_k$ is dependent on different retransmission trigger, 1 for InterCnt exceeding CntThres and 0 for packet interarrival timeout.
To determine segment-level CntThres, the expected packet disorder length is estimated through sampling the InterCnt of out-of-order packets at each retransmission node of its own CR segment. When a new out-of-order packet is observed, an exponentially weighted moving average and its mean deviation of a disordered packet length are iteratively updated, upon which the CntThres in the expected packet list is updated.

\emph{Retransmission Timeout} -- After packet loss is detected, both the transmitted RR and RD packets can be lost during the retransmission phase. Therefore, a retransmission node should be able to detect the retransmitted packet loss and resend the retransmission request. To determine a threshold for retransmission timeout, each packet retransmission delay (i.e., retransmission RTT) is sampled to estimate an expected retransmission RTT, given by \cite{paxson2011computing}
\begin{equation}
\setlength{\abovedisplayskip}{3pt} 
\setlength{\belowdisplayskip}{3pt}
\begin{split}
R_v(k+1) &= (1-\xi)R_v(k)+\xi R(k+1)\\
R_m(k+1) &= (1-\delta)R_m(k)+\delta |R(k+1)-R_v(k)|\\
R_{th}(k+1) &= R_v(k+1) + \phi R_m(k+1),\, k \in \mathbf{Z}
\label{equ:RTO_R}
\end{split}
\end{equation}
where $R(k)$ is the $k\,$th retransmission RTT sample, $R_v(k)$ is an expected retransmission RTT calculated from $k$ samples, $R_m(k)$ is a mean deviation from $k$ retransmission RTT samples, $R_{th}(k+1)$ is the updated threshold for retransmission timeout (i.e., expected retransmission RTT) after the $(k+1)\,$th retransmission RTT sample, $R_v(0) = R(1)$, $R_m(0) = 0$, $\xi = 0.125 $, $\delta = 0.25$, and $\phi = 4$.
	
\subsubsection{Retransmission Nodes Triggering RR Packets}
\label{sec:SDP_RP_RR}
After a retransmission node detects packet loss, it sends an RR packet to the preceding caching node, requesting the retransmission of the lost packet(s). If the retransmission is triggered by a packet-level interarrival counter threshold, the node also needs to generate and send an RI packet downstream to the following retransmission nodes at the same time.
Main fields of the RR packet are similar to those of RI packet, expect that AddL is replaced by the Timestamp field, which records the time of sending an RR packet. The Flag field also indicates how an RR packet is triggered. For RR packets triggered by interarrival counter threshold (C) and retransmission timeout (R), StartSeq and EndSeq specify that the packets with the sequence numbers lying in between StartSeq and EndSeq are expected to be retransmitted; For RR packets triggered by interarrival timeout (T) where EndSeq fields are unknown, we use StartSeq and StartNum to locate each specific expected packet to be retransmitted.
\subsubsection{Caching Nodes Retransmitting RD Packets}
\label{sec:SDP_RP_RP}
When a caching node receives an RR packet, a range of sequence numbers for the requested packets can be obtained from the RR packet. The caching node searches in its data caching buffer for the requested packets. If the requested packets are successfully cached and are not triggered by the same condition (i.e., interarrival timeout or interarrival counter threshold), the RD packets are sent out. Similar to RR packets, each RD packet includes the timestamp fields, for the retransmission timeout detection in case of retransmitted packet loss, and includes the values of StartSeq and StartNum from its received RR packet and the requested data payload; If the requested packets are not found in current caching node, the RR packet is forwarded to each preceding caching node consecutively until the packets are found in the current CR segment.

\vspace{-0.1cm}
\section{Simulation Results}
\label{sec:SDP_S}
\begin{figure}[!htb]
	\vspace{-0.4cm}  
	\setlength{\belowcaptionskip}{-0.4cm}   
	\centering
	{\includegraphics[scale=0.25]{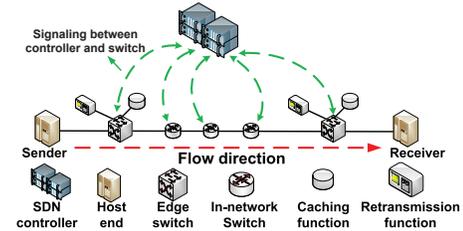}}
	\caption{Simulation topology.}\label{fig:single_path_retransmission_topo_simulation}
\end{figure}

In this section, simulation results are presented to demonstrate the effectiveness of the proposed protocol. 
Since SDTP is proposed to achieve reliable end-to-end packet transmission, similar as TCP, the simulation results are compared with TCP to show the effectiveness achieved by applying SDTP with SDN controller.
 We use two separated virtual machines to simulate the control plane and the data plane, respectively, of an SDT-based embedded virtual network topology, where we have an SDN controller, two end hosts and five switches as shown in Fig.~\ref{fig:single_path_retransmission_topo_simulation}. Each virtual machine utilizes a 4 GB physical memory and an Intel Core i7-4770HQ CPU at 2.20GHz with a dual-core processor. The SDN control plane is implemented by the Ryu framework \cite{Rub1}. The network elements including end hosts, switches, and transmission links are emulated by Mininet \cite{RN88}. The switches in the simulation are Open vSwitches \cite{Rub2}. We choose OpenFlow Version 1.3.0 \cite{RN89} to implement the SDN southbound interface.
The link delay between consecutive switches is set to $5\,$ms, and the link between an end host and an edge switch has $20\,$ms delay. The E2E packet loss rate ranges from $0\%$ to $5\%$. For each sending-receiving E2E transmission connection, a sending host sends every packet per $15\,$ms. For the control plane, there is a dedicated link used for signaling exchanges between the controller and each switch. The SDN control efficiency depends on the controller processing capacity and the distances between the controller and each switch. In our simulations, we set control delay (i.e., the time duration from the instant that the sending edge switch sends a control message to the instant that the receiving edge switch receives the control message) as a variable to indicate different levels of control efficiency. 

\subsection{Connection Establishment}
\begin{figure}[!htb]
	\vspace{-0.4cm}  
	\setlength{\belowcaptionskip}{-0.6cm}   
	\centering
	{\includegraphics[scale=0.2]{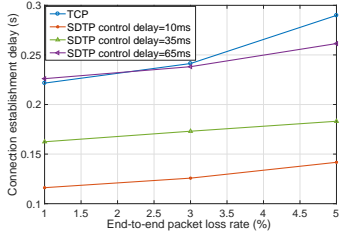}}
	\caption{Connection establishment delay of SDTP and TCP.}\label{fig:experiment_conn_delay_1}
\end{figure}

Fig.~\ref{fig:experiment_conn_delay_1} shows the relationship of E2E connection establishment delay versus packet loss rate. The connection delay increases with the control delay in SDTP. With a small control delay ($10\,$ms), the SDTP connection delay is much less than that of TCP. When the control delay becomes large ($65\,$ms), SDTP has a larger connection delay than TCP if the E2E packet loss rate is small, since few connection signaling packets are lost and retransmitted. With an increased packet loss rate, the two-way handshake to establish a connection for SDTP avoids ACK packet loss, reduces the probability of restarting the connection establishment, and thus achieves low connection delay as compared to TCP.
\vspace{-0.1cm}
\subsection{Average End-to-End Packet Delay}
\begin{figure}[!htb]
	\vspace{-0.4cm}  
	\setlength{\belowcaptionskip}{-0.1cm}   
	\centering
	{\includegraphics[scale=0.2]{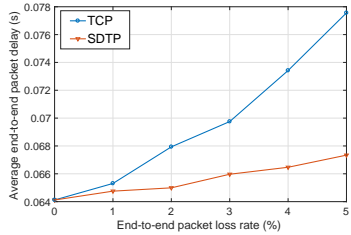}}
	\caption{Average E2E packet delay of SDTP and TCP.}\label{fig:delay_average}
\end{figure}

\begin{figure}[!htb]
	\vspace{-0.3cm}  
	\setlength{\belowcaptionskip}{-0.3cm}   
	\centering
	{\includegraphics[scale=0.58]{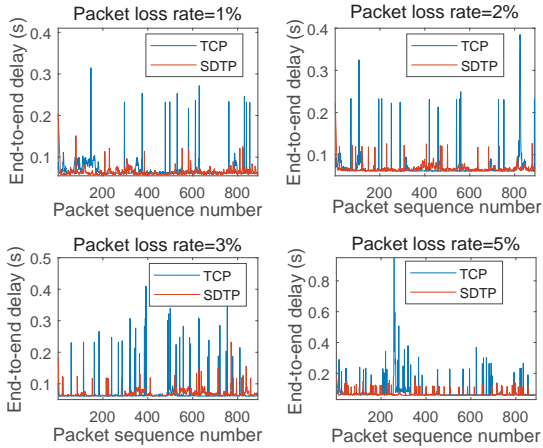}}
	\caption{E2E packet delay with different packet loss rates.}\label{fig:delay_seqnum}
\end{figure}

In Fig.~\ref{fig:delay_average}, we evaluate the average E2E packet delay (i.e., the duration from the time a data packet is sent from a sending host till the time instant it is successfully received by a receiving host) with different packet loss rates. We also evaluate the E2E packet delay distribution for the proposed protocol and TCP. One thousand continuous packet samples are performed and the results are shown in Fig.~\ref{fig:delay_seqnum}. For different packet loss rates, TCP incurs higher delay jitter, which indicates that the SDTP performs more stable data transmissions. As the packet loss rate increases, both packet delay and delay jitter of TCP are enlarged much more than those in SDTP, since the lost packets of TCP can only be detected and retransmitted by the source node. However, for the SDTP, the lost packets can be detected earlier by the in-path retransmission nodes and be retransmitted faster by the in-path caching nodes.
\vspace{-0.3cm}
\vspace{-0.1cm}
\section{Conclusion}
\label{sec:Conclusion_future_work}
In this paper, we present a novel SDN-based transmission protocol (SDTP) with simplified connection establishment and enhanced data transmission efficiency. Specifically, the SDTP replaces the three-way handshake mechanism in TCP with a new two-way connection establishment procedure, which offloads the signaling overhead from the data plane to the control plane, and reduces the amount of signaling exchanges overhead. For data transmission, a new header format is designed to support efficient data transmission with in-path packet retransmission.
Based on the new SDTP packet header format, a novel in-path receiver-based packet loss detection and caching-based packet retransmission scheme is proposed to achieve fast lost packet detection and recovery with low signaling overhead.
Extensive simulation results are provided to demonstrate the advantages of the proposed SDTP over the conventional TCP. For future work, we will optimize the caching/retransmission function placement to minimize the E2E packet delay, and also investigate caching-based in-network congestion control.


\bibliographystyle{IEEEtran}
\bibliography{Reference}

\end{document}